\documentstyle[prl,aps,epsf,preprint]{revtex}

\tighten
\newcommand{\be}{\begin{equation}}              
\newcommand{\ee}{\end{equation}}                
\newcommand{\bea}{\begin{eqnarray}}             
\newcommand{\eea}{\end{eqnarray}}
\newcommand{\nn}{\nonumber}
\newcommand{\bm}[1]{\mbox{\boldmath${#1}$}}

\tighten                                

\begin{document}

%
\title{Magnus and  Iordanskii Forces in Superfluids}
\author{C. Wexler}
\address{Department of Physics, Box 351560, 
         University of Washington, 
         Seattle, WA 98195-1560}
\date{April 1997}
\maketitle

\begin{abstract}
The total transverse force acting on a quantized vortex in a
superfluid is a problem that has eluded a complete understanding for
more than three decades.
In this letter I propose a remarkably simple argument, somewhat
reminiscent of Laughlin's beautiful argument for the quantization of
conductance in the quantum Hall effect\cite{laughlin}, to define the
{\em superfluid } velocity part of the transverse force.  This term is
found to be $-\rho_s \; \bm{\kappa}_s \times \bm{v}_s$. Although this
result does not seem to be overly controversial, this thermodynamic
argument based only on macroscopic properties of the superfluid does
offer a robust derivation.
A recent publication by Thouless, Ao and Niu\cite{tan96} has
demonstrated that the {\em vortex } velocity part of the transverse
force in a homogeneous neutral superfluid is given by the usual form
$\rho_s \; \bm{\kappa}_s \times \bm{v}_V$\cite{per_length}.
A combination of these two independent results and the required
Galilean invariance yields that there cannot be any transverse force
proportional to the normal fluid velocity, in apparent conflict with
Iordanskii's theory of the transverse force due to phonon scattering
by the vortex\cite{iordanskii}.
\end{abstract}

%

\section{Introduction}

The importance of quantized vortices in superfluids has been
recognized since Onsager first put forth the idea of quantization of
circulation almost 50 years ago\cite{vort_intro,feynman55}.
A vortex moving in a superfluid experiences a force transverse to its
velocity which is equivalent to the Magnus or Kutta-Joukowski
hydrodynamic lift present in classical hydrodynamics\cite{lamb32},
which is generally written\cite{per_length}

\be
\label{eq:KJ_force}
\bm{F} = \rho \; \bm{\kappa} \times ( \bm{v}_V - \bm{v}_{\text{fluid}} ).
\ee

However, there is no consensus on the problem of generalizing the
Magnus force to the superfluid case, and various inequivalent
expressions for the relevant forces
abound\cite{tan96,barenghi83,qvih2,volovik95,demircan95,sonin97,wexler96}.

This letter deals specifically with the calculation of the {\em
superfluid velocity part of the Magnus force}, that is the transverse
force that depends on the superfluid velocity $\bm{v}_s$.  It is shown
that this force is given by

\be
\label{eq:vs_dep_force}
\bm{F}_s = - \rho_s \; \bm{\kappa}_s \times \bm{v}_s . 
\ee

Recently Thouless, Ao and Niu\cite{tan96} (TAN) have convincingly
argued for a universal {\em vortex velocity part of the Magnus force}
given by

\be
\label{eq:TAN}
\bm{F}_V = \rho_s \; \bm{\kappa}_s \times \bm{v}_V 
\ee

\noindent
for uniform neutral superfluids; and Geller, Wexler and Thouless have
generalized TAN's results to charged systems in the presence of a periodic
potential\cite{geller}.
The natural assumption by TAN that the normal fluid does not circulate
around the vortex $(\kappa_n = \oint \bm{v}_n \cdot d \bm{l} = 0)$ has
been confirmed by a recent calculation\cite{wexler96} in the
thermodynamic limit (mean free path of excitations much smaller than
system size).  There is an important point to notice: TAN deals only
with the part of the transverse force that depends on the {\em vortex}
velocity, while no statement is made regarding the other parts of this
force that depend on the normal and superfluid velocities.

The confusion in the topic of writing the various parts of the Magnus
force is widespread. Part of it arises from different interpretations
on the role played by excitations, and whether or not these are
scattered asymmetrically by the vortex\cite{iordanskii} leading to a
transverse force (the Iordanskii force) proportional to the normal
fluid density $\rho_n$ and either the relative velocity 
$(\bm{v}_n - \bm{v}_V)$ or $(\bm{v}_n - \bm{v}_s)$.

The results presented in this letter, combined with TAN and Galilean
invariance, are incompatible with the existence of a transverse force
proportional to the normal fluid velocity $\bm{v}_n$.

I must note that this letter does not deal with the determination of
additional dissipative terms (namely the {\em longitudinal} forces),
which are negligible under the conditions of the present work. This is
an interesting subject as well, but the arguments presented only
determine the transverse forces.

In sec.~\ref{sec:gal_inv} I write down the most general transverse
force, linear in the velocities, that is compatible with Galilean
invariance. Section \ref{sec:free_ener_and_force} is the main section
of the letter, where the the superfluid velocity part of the Magnus
force is calculated. In sec.~\ref{sec:the_rest_of_it} I write the
final form of the transverse force by combining the results of this
letter and TAN, plus Galilean invariance. I also discuss the diverse
results obtained by other authors and their assumptions.

\section{Galilean Invariant Transverse Force}
\label{sec:gal_inv}

In a homogeneous superfluid the forces acting on a vortex must be
expressed in terms of velocity differences only. Consider a vortex
moving with velocity $\bm{v}_V$ in a superfluid where the superfluid
component has an asymptotic velocity $\bm{v}_s$ and the normal
component $\bm{v}_n$. The most general {\em Galilean invariant
transverse force} can be written as

\be
\label{eq:gal_inv_force}
\bm{F} = A \; \bm{\hat{\kappa}} \times ( \bm{v}_V - \bm{v}_s ) +
         B \; \bm{\hat{\kappa}} \times ( \bm{v}_V - \bm{v}_n ) ,
\ee

\noindent
where $A$ and $B$ are constants to be determined and
$\bm{\hat{\kappa}}$ is a unit vector pointing in the direction of the
vortex line. Our task is the determination of these unknown
coefficients. It is customary to divide this expression into separate
terms, each involving one particular velocity and denote them
accordingly: the {\em vortex velocity} part of the Magnus force 
$(A+B) \; \bm{\hat{\kappa}} \times \bm{v}_V$, the {\em superfluid velocity}
part $(- A) \; \bm{\hat{\kappa}} \times \bm{v}_s$, and the {\em normal
fluid velocity} part $(- B) \; \bm{\hat{\kappa}} \times \bm{v}_n$. 
Knowledge of two of these forces completely determines the
third. In the following section I determine the coefficient $A$ by
calculating the superfluid velocity part of the Magnus force.

\section{Free Energy and Force}
\label{sec:free_ener_and_force}

Here I wish to present a very simple {\em gedanken} experiment, whose
outcome will determine the coefficient $A$ as mentioned above. This
argument has some parallels to Laughlin's own thought experiment
relating to the quantization of Hall conductance in the quantum Hall
effect\cite{laughlin}.

Consider a neutral superfluid trapped inside a toroid like the one
shown in the figure. For simplicity assume the toroid to have a
roughly uniform section and that the circumference $L_x$ is much
bigger than $L_y$. This makes the superfluid velocity $v_s$
approximately uniform, which is what is actually desired for a
definition of the {\em superfluid velocity} part of the Magnus
force. I must remark that the argument is more general, and these
assumption are merely necessary to keep the argument clean and simple.

Assume that in the initial state $N \gg 1$ quanta of circulation are
trapped in the toroid so that the superfluid velocity is given by 
$v_s = N h/m L_x$. Under this condition the normal fluid is pinned to
the container and $v_n$ is zero
(in fact, this is how the normal density
$\rho_n$ is normally {\em defined}\cite{landau_rho_n}). At some
initial time a vortex is created at the outer edge and {\em slowly}
dragged towards the center of the ring by some means\cite{tan96},
where it is annihilated at a later time $t = \tau \rightarrow \infty$. 
The final state corresponds to a trapped circulation $(N+1) \: h/m$, 
while the normal fluid will still be at rest. By performing
this process very slowly, dissipative effects are negligible.

While transporting the vortex across the ring, one needs to perform
work on the system. In terms of eq. (\ref{eq:gal_inv_force}), and
given the fact that $v_V = v_n = 0$, the work is given by integrating
the force along the displacement $d\bm{r}$ of each vortex segment
$d\bm{l}$:

\bea
\label{eq:work}
W = - &&\int  \; A \; (d\bm{l} \times \bm{v}_s) \cdot d\bm{r} 
  = A \; \int (d\bm{l} \times d\bm{r}) \cdot \bm{v}_s \nn \\
  && = A \; \int d\bm{S} \cdot \bm{v}_s 
  = A \; (L_y L_z) \; v_s .
\eea

For {\em isolated} systems the change in energy corresponds to the
amount of work. The argument is straightforward but it is much simpler
to consider an {\em isothermal} process. The amount of work performed
then corresponds to the variation of the Helmholtz {\em free} energy
${\cal A} = {\cal E} - TS $ \cite{chandler}. The free energy can be
expressed in terms of the energy of the ground state plus the free
energy of excitations:

\be
\label{eq:free_energy_decomposed}
{\cal A} = E_{\text{g.s.}} + {\cal A}_{\text{excit}} .  
\ee

One will be interested in the variation of the free energy due to a
variation of the superfluid velocity $v_s$, therefore one needs only
to consider the $v_s$ dependent portions of ${\cal A}$. The relevant
ground state energy is

\be
\label{eq:energy_gs}
E_{\text{g.s.}} = (L_x L_y L_z) \rho \; v_s^2/2 ,
\ee

\noindent
and the excitation free energy is given by the standard expression

\be
\label{eq:free_energy_excitations}
{\cal A}_{\text{excit}} = (k_B T)  \sum_{\text{modes}} 
        \log \left( 1  -  e^{- \epsilon / k_B T} 
                \right) ,
\ee

\noindent
where the excitation energies $\epsilon$ are those in the 
``rest frame'', and are therefore Doppler shifted by the superflow:

\be
\label{eq:doppler_shift_energies}
\epsilon = \hbar \omega ({k}) + v_s \hbar k_x .
\ee

To second order in the superfluid velocity the free energy is given by

\bea
{\cal A}_{\text{excit}} - &&{\cal A}_{\text{excit}}(v_s=0)
\; = \;   \frac{v_s^2}{2}
       \left. \frac{\partial^2 {\cal A}_{\text{excit}}}{\partial v_s^2} 
      \right|_{v_s=0} \nn \\
&& =  - \frac{\hbar^2 v_s^2}{2 k_B T}  \sum_{\text{modes}}
        k_x^2 \frac{e^{\hbar \omega (k)/k_B T}}
               {(e^{\hbar \omega (k)/k_B T} - 1 )^2}  \nn \\
&& = - (L_x L_y L_z) \; \rho_n \; \frac{v_s^2}{2} ,
\eea

\noindent
where the last equality follows from the usual Landau derivation of the
normal density\cite{landau_rho_n}. 

The total change in Helmholtz free energy for variations in the
superfluid velocity can be written as\cite{therm_def}

\bea
\label{eq:var_A_vs}
\Delta {\cal A} && = (L_x L_y L_z) \: \frac{\rho - \rho_n}{2} \: 
        \Delta (v_s^2)  \nn \\
&& = (L_x L_y L_z) \: \frac{\rho_s}{2} \: \Delta (v_s^2) 
 = (L_y L_z) \: \rho_s \: v_s \frac{h}{m} ,
\eea

Where the last equality corresponds to the change $\Delta v_s = h/m L_x$ 
due to the motion of the vortex across the ring.

By equating the work performed on the system (eq. \ref{eq:work}) and
the variation of free energy (eq. \ref{eq:var_A_vs}) the unknown
coefficient $A$ in the general expression for the transverse force
(\ref{eq:gal_inv_force}) can be determined:

\be
\label{eq:A_determined}
A = \rho_s \; \frac{h}{m} .
\ee

\section{Total Transverse Force and Conclusions}
\label{sec:the_rest_of_it}

Having calculated the {\em superfluid} velocity part of the transverse
force, there is the need to obtain one more component of it to
completely determine the transverse force (\ref{eq:gal_inv_force}).

By considering TAN's result\cite{tan96,wexler96}, as written in
eq. (\ref{eq:TAN}), it is clear that the coefficient of the 
{\em vortex} velocity $(A+B)$ is given by

\be
\label{eq:A_plus_B}
(A + B) = \rho_s \; \frac{h}{m} ,
\ee

\noindent
and this, along with the result for $A$ calculated in the previous
section, yields unambiguously $B = 0$, meaning that there is no
transverse force depending on the {\em normal fluid} velocity. The
total {\em transverse} force acting on a vortex would be simply
written

\be
\label{eq:tot_trans_force}
\bm{F} = \rho_s \; \frac{h}{m} \; \bm{\hat{\bm{z}}} \times 
        ( \bm{v}_V - \bm{v}_s ) ,
\ee

\noindent 
with the transverse Iordanskii force vanishing exactly.

This is in general agreement with some direct calculations of the
normal fluid velocity part of the transverse force based on the
scattering of excitations by the vortex\cite{demircan95,wexler97}.
This calculations also show that the coefficient of the normal fluid
velocity either vanishes or is much smaller than previously thought,
in apparent conflict with Iordanskii's theory of the transverse force
on a vortex\cite{iordanskii,sonin97}.

It is interesting to note that while this letter, in combination with
TAN's result, yields an {\em exactly} vanishing Iordanskii force, the
``direct'' calculations mentioned above either give a non-zero
result\cite{iordanskii,sonin97} or can only hint that it is
small\cite{demircan95,wexler96}.

One must emphasize some differences in the assumptions about the
asymptotic flow of excitations far away from the vortex. The authors
calculating directly the Iordanskii force from the excitation
scattering assume a homogeneous distribution of non-interacting
phonons\cite{iordanskii,demircan95,sonin97,wexler97}, while the
calculations used along this letter and references \onlinecite{tan96}
and \onlinecite{wexler96} include the effect of the vortex in the
excitation distribution. This can explain some of the apparent
discrepancies: a careful calculation of the Iordanskii force must
include {\em both} the effects of the scattering of these excitations
{\em and} the perturbation of the distribution functions by the
vortex. Similar effects have been long known in the calculation of the
viscous drag of a moving object in a fluid, namely the Stokes
problem\cite{stokes_problem}. It may be possible to obtain
independently the exact cancelation of the transverse Iordanskii force
by including all the effects described in this paragraph. While this
would be certainly desirable, it goes beyond the scope of this letter.

\acknowledgements

I wish to thank David Thouless,  Michael Geller, Jung Hoon Han, Greg
Dash and John Rehr for numerous helpful discussions. 
This work was supported by the NSF grant DMR-9528345.




\begin{figure}
        \begin{center}
        \leavevmode
        \epsfbox{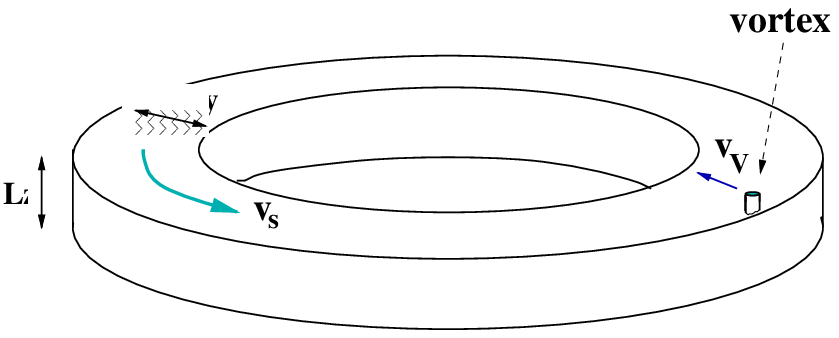}   
        \caption{ \label{fig:gedankevortex}
                {\em Gedanken} experiment: a vortex is
                adiabatically created at the outer edge of a toroid
                with circumference $L_x$, transported
                across the channel and annihilated at the inner edge
                thus increasing the total circulation around the
                toroid by one quantum of circulation $h/m$.}
        \end{center}
\end{figure}

\end{document}